\newcommand{\be}{\begin{equation}}
\newcommand{\ee}{\end{equation}}
\newcommand{\bea}{\begin{eqnarray}}
\newcommand{\eea}{\end{eqnarray}}
\documentstyle[12pt]{article}
\begin{document}
\thispagestyle{empty}
\vspace{20cm}

\begin{center}

{\large\bf INTEGRABLE EXTENSIONS OF N=2 SUPERSYMMETRIC KdV HIERARCHY 
ASSOCIATED WITH THE NONUNIQUENESS OF THE ROOTS OF THE 
LAX OPERATOR}
\vspace{1.5cm} \\
 Z. Popowicz \footnote{E-mail: ziemek@ift.uni.wroc.pl}
\vspace{1.0cm} \\
{Institute of Theoretical Physics, University of Wroclaw \\
pl.M.Borna 9, 50-204 Wroclaw, Poland}  \vspace{1.5cm}
\end{center}

\begin{center}
{\bf Abstract.}
\end{center}
We preesent a new supersymmetric integrable extensions of the a=4,N=2 
KdV 
hierarchy. The  root of the supersymmetric Lax 
operator of the KdV equation is generalized, by including additional  
fields. This generalized root generate new hierarchy of integrable 
equations, for which we investigate the hamiltonian structure. 
In special case our system describes the 
interaction of the KdV equation with the two MKdV equations.

 \newpage

\section{Introduction}
The Korteweg -de Vries (KdV) equation, which has been extensively 
studied by 
mathematicans as well as physicist [1-3] in the last 30 years, is 
probably 
the most popular soliton equation. On the other side, various 
different 
generalizations of the soliton equation have recently been proposed 
as the 
Kadomtsev - Petviashvilli and Gelfand - Dickey hierarchies and 
supersymmetrization [4-11]. The motivation for studing these are 
diverse. 
In the supersymmetric generalization, one expects that, in the so 
called 
bosonic sector of the supersymmetry (SUSY) a new class of the 
integrable 
models may appear. 

From  the soliton point of view we can distinguish two important 
classes of 
the supersymmetric equations: the non-extended $(N=1)$ and extended 
$(N>1)$
cases. Considerations of the extended case may imply new bosonic 
equations whose properties need further investigation. This may be 
viewed as 
a bonus, but this extended case is in no way more fundamental than 
the 
non-extended one.

In order to get a SUSY theory we have to add to a system of $k$ 
bosonic 
equations $kN$ fermions and $k(N-1)$ boson fields $(k=1,2... 
N=1,2..)$ in 
such a way that the final theory becomes SUSY invariant. 
Interestingly 
enough, it appeared that during the supersymmetrizations, some 
typical SUSY 
effects (compared with the classical theory) occured. For example, 
the Lax 
operator for the one of the extended $(N=2)$ supersymmetric extension 
of the 
KdV equation (a=4) possesses nonunique roots [12]. The supersymmetric 
Boussinesq [13] equation does not reduce to the classical Boussinesq 
equation. The N=1 supersymmetric KdV equation possesses the non-local 
conservation laws [14]. These effects rely strongly on the 
descriptions of 
the generalized systems of equations which we would like to 
supersymmetrize.

Until now supersymmetric KdV equation have been
constructed for $N=1,2,3$ and $4$ [4,8,15-18] based on their relation 
to the 
superconformal algebras. For extended $N=2$ supersymetric case it 
appeared 
that it is possible to construct three different integrable 
extensions of the KdV eqution [8,19,20]. All these extensions have 
Lax pair 
representations. On the other side a new variety of generalizations 
of all 
different supersymmetric extensions of the KdV equation have been 
proposed [10,21-23] recently. In order to obtain such extensions the 
Lax 
operator is modified by including, in a proper way, the additionals 
fileds. These Lax operators produce new integrable hierarchy of 
equations. 
However such procedure is technically complicated.

For the supersymmetric $N=2,a=4$ KdV equation we have additional 
possibility 
of the generalizations of this equation.
Indeed, we can generalize one of the root of the Lax operator, by 
including 
to it  additional fields. As the result the Lax operator is also 
generalized. It simplify the process of gehneralization of the Lax 
operator.
In this paper we describe such construction in 
more details.  As a result we obtain new extensions of the $N=2,a=4$ 
supersymmetric KdV equation. This extensions constitute the 
hamiltoninan 
system. We investigate also the  reduction of our 
generalized supersymmetric equation. Finally let us mention that our 
generalization  is different then this proposed by Ivanov and Krivons 
[21].

\section{Notation}
The basic objects in the supersymmetric analysis are superfileds and 
the 
supersymmetric derivatives. The Taylor expansion of the superfield 
with 
respect to the $\theta$ is \be
\phi(x,\theta_{1},\theta_{2})=w(x)+\theta_{1}\xi_{1}+\theta_{2}\xi_{2}
\theta_{2}\theta_{2}u,
\ee
where the fields $w,u$ are to be interpreted as the boson (fermion) 
fields 
for the superboson (superfermion) field while $\xi_{1},\xi_{2}$, as 
the 
fermion (boson) for the superboson (superfermions) respectively. The
superderivatives are defined as
\bea
D_{1}=\partial_{\theta_{1}} + \theta_{1}\partial,\\ \
D_{2}=\partial_{\theta_{2}} + \theta_{2}\partial,
\eea
and satisfy $D_{1}^{2}=D_{2}^{2}=\partial$ and $ D_{1}D_{2} + 
D_{2}D_{1} =0$.

Below we shall use the following notation: $(D_{i}F)$ denotes the 
outcome
of the action of the superderivative on the superfield F, while 
$D_{i} F$ 
denotes the action itself of the superderivative on the superfiled F.

\section{Supersymmetric N=2 KdV equation}

This equation could be written down in the following one-parameter 
family
of super - Hamiltonian evolution equations
\bea
\phi_{t}& & =  -\phi_{xxx}+3(\phi D_{1}D_{2}\phi )_{x} +
\frac{1}{2}(a-1)(D_{1}D_{2}\phi^{2})_{x}+3a\phi^{2} \phi_{x} \ \\
&& = (D_{1}D_{2}\partial +2\partial\phi+2\phi \partial -D_{1}\phi 
D_{1}
-D_{2}\phi D_{2})\frac{\delta}{\delta\phi}\int\frac{1}{2}(\phi 
D_{1}D_{2}\phi+\frac{a}{3}\phi^{3})dX
\eea
where $dX=dxd\theta_{1}\theta_{2}$ and  $a$ is an arbitrary parameter 
and $\phi$ is a superboson.

The operator 
\be
P_{2}(\phi) := D_{1}D_{2}\partial + 2\partial \phi + 2\phi \partial - 
    D_{1}\phi D_{1} -D_{2}\phi D_{2}
\ee
in (5) is the hamiltonian operator stemming from the $N=2$ extension 
of the 
Virasoro algebra.

Although for arbitrary values of $a$ a Miura transformation for (4) 
had been given in [8], only for three cases $(a=-2,4,1)$ Lax 
formulations 
had been found [8,20]. They are given by
\bea
a=1 : && {~}{~}  L:= \partial + \partial^{-1}D_{1}D_{2}\phi \ \\
a=-2: && {~}{~}  L:=\partial^{2}+ D_{1}\phi D_{2} - D_{2}\phi D_{1}\ 
\\
a=4 : && {~}{~}  L:=\partial^{2}-(D_{1}D_{2}\phi) -\phi^{2} 
+(D_{2}\phi)D_{1} -(D_{1}\phi)D_{2} -2\phi D_{1}D_{2}
\eea
For $a=-2,4$ the equation (4) is equivalent to [8]
\be
\frac{d}{dt}L = -4\Big[ (L^\frac{3}{2})_{+},L\Big]
\ee 

where ${+}$ denotes the (super) differential part of the 
operator. 

For $a=1$ the equation (4) is equivalent to [20]
\be
\frac{d}{dt}L = \Big[ (L^{3})_{\geq 1},L \Big]
\ee
where ${\geq 1}$ denotes purely (super) differential part of the 
operator.

The fractional powers of L can be obtained from the square roots of 
the Lax 
operators defined by the following expansion into (inverse) powers of 
the 
differential operator
\be
L^{\frac{1}{2}}= \partial + \sum_{k=1}^{\infty}(a_{k} +b_{k}D_{1} 
+c_{k}D_{2} + d_{k}D_{1}D_{2})\partial^{-k} ,
\ee
where $a_{k},d_{k}$ are superbosons while $b_{k},c_{k}$ are 
superfermions.

However such square root can be defined also in a different manner 
\be 
L^{\frac{1}{2}} =i \Big( D_{1}D_{2} + a_{0} + 
\sum_{k=1}^{\infty}(aa_{k} + 
bb_{k}D_{1}+cc_{k}D_{2}+dd_{k}D_{1}D_{2})\partial^{-k}\Big ).
\ee
where $aa_{k},dd_{k}$ are superbosons while $bb_{k},cc_{k}$ are 
superfermions.

Indeed the Lax operator (9) for the $N=2,a=4$ supersymmetric 
KdV equation can be written down as
\be
\hat { L}= D_{1}D_{2} +\phi.
\ee
Due to it also Lax equations of the type 
\be
\frac{d}{dt_{n}}\hat L = \Big[ (\hat L^{n} L^{\frac{1}{2}})_{+},\hat 
L\Big ]
\ee
can be considered in addition to (10). Furthemore from the Lax 
operator 
$\hat L$ additional conserved quantitites for the equation of motion 
are 
given by the residues of the pseudo-differential operators 
$L^{k+\frac{1}{2}}\hat L $. These correspond to the additional 
Hamiltonian 
functions $H_{2k}$ which reduce to 0 when passing to the $N=1$ case.

\section{The generalization of the supersymmetric a=4 KdV equation}

The formula (13) suggests to consider more general form of the root 
of the 
Lax operator as  (14). Therefore we consider two different 
generalizations of the operator (14) : 
\be
L_{b} = D_{1}D_{2} + \phi + h\partial^{-1}D_{1}D_{2}g - 
g\partial^{-1}D_{1}D_{2}h 
\ee
\be
L_{f} = D_{1}D_{2} +\phi + h\partial^{-1}D_{1}D_{2}g + 
g\partial^{-1}D_{1}D_{2}h
\ee
where $h,g$ are the superbosons for $L_{b}$ while they are 
superfermions for 
$L_{f}$.

Now it is rather technical problem to construct the Lax operator as 
the 
second power of $L_{f}$ or $L_{b}$. 
Next we can easily compute second root of such constructed Lax 
operator, 
using the formula (12) and finally to construct the whole hierarchy 
of 
integrable equations utilizing the formula (15). 
Explicitely the second flow for the superbosonic case is

\bea
\phi_{t} &=& \Big ( -D_{1}D_{2}\phi -2\phi^{2} - 2(D_{2}gD_{2}h) -
2(D_{1}gD_{1}h) +2gh_{x} -2g_{x}h\Big )/2, \ \\
h_{t} &=& -\Big ( 2D_{1}D_{2}h_{x} + (D_{1}\phi D_{1}h) +(D_{2}\phi 
D_{2}h) +
\phi_{x}h +4(D_{2}hD_{1}h)g \cr
&& +2(D_{1}gD_{2}h)h -2(D_{2}gD_{1}h)h +2(D_{1}D_{2}g)h^{2} 
-4(D_{1}D_{2}h)gh +4\phi h_{x}\Big )/2 \ \\
g_{t} &=& -\Big ( 2D_{1}D_{2}g_{x} +(D_{1}\phi D_{1}g) +(D_{2}\phi 
D_{2}h) +
\phi_{x}g -4(D_{2}gD_{1}g)h) \cr
&& -2(D_{1}gD_{2}h)g +2(D_{2}gD_{1}h)g + 2(D_{1}D_{2}g)gh 
-2(D_{1}D_{2}h)g^{2} +4\phi g_{x}\Big )/2,
\eea
while for the superfermionic case is 

\bea
\phi_{t} &=& \Big (-(D_{1}D_{2}\phi) -2\phi^{2}+ 2g_{x}h - 2gh_{x} 
-2(D_{1}gD_{1}h) -2(D_{2}gD_{2}h)\Big )_{x}/2, \ \\
h_{t} &=& \Big (-2D_{1}D_{2}h_{x} -4h_{x}\phi -(D_{1}\phi D_{1}h) 
-(D_{2}\phi D_{2}h) -2g(D_{1}D_{2}h)h \cr
&& -h\phi_{x} + 2h(D_{1}gD_{2}h)  -2h(D_{2}gD_{1}h \Big )/2, \ \\
g_{t} &=& \Big ( -2D_{1}D_{2}g_{x} -4g_{x}\phi -(D_{1}\phi D_{1}g) -
(D_{2}\phi D_{2}g) -2(D_{1}D_{2}g)gh \cr
&& -g\phi_{x}- 2g(D_{1}gD_{2}h) + 2g(D_{2}gD_{1}h \Big )/2.
\eea

The conserved charges for our systems are given by the residues of 
the 
pseudo-di\-ffe\-ren\-tial opertaors $L^{k+\frac{1}{2}}\hat L$. 
Explicitely the first three charges are:

for the superbosonic case
\bea
H_{1} &=& \int dX \phi, \ \\
H_{2} &=& \int dX \Big ( 4g_{x}h + \phi^{2} \Big ), \ \\
H_{3} &=& \int dX \Big ( \phi (D_{1}D_{2}\phi ) +\frac{4}{3}\phi^{3} 
+6h(D_{1}D_{2}g_{x}) + 3(D_{1}g)(D_{1}h)\phi +\cr
&& 3(D_{2}g)(D_{1}g)\phi + 3g_{x}h\phi - 3gh_{x}\phi - 
3(D_{1}g)(D_{2}h)gh -\cr
&& 3(D_{2}g)(D_{1}g)h^{2} +3(D_{2}g)(D_{1}h)hg - 
3(D_{2}h)(D_{1}h)g^{2} \Big 
)/12.
\eea

while for the superfermionic case 
\bea 
H_{1} &=& \int dX \phi ,\ \\
H_{2} &=& \int dX \Big (-4g_{x}h + \phi^{2}\Big ), \ \\
H_{3} &=& \int dX  \Big ( \phi (D_{1}D_{2}\phi ) + 
\frac{4}{3}\phi^{3} 
-6(D_{1}D_{2}g_{x})h - 3g_{x}h\phi + 3gh_{x}\phi \cr
&& -3gh(D_{1}g)(D_{2}h) + 3gh(D_{2}g)(D_{1}h) + 3\phi 
(D_{1}g)(D_{1}h) +\cr
&& 3\phi (D_{2}g)(D_{2}h \Big )/12.
\eea

The systems of equations (18-20) and (21-23)  are hamiltonians 
systems with 
the following Hamiltonian operator:
\newpage

for the superbosonic case
\be
P =\pmatrix{ \partial & 0 & 0 \cr
        0 & 0 & -\frac{1}{2} \cr
        0 & \frac{1}{2} & 0},
\ee

and for the superfermionic case
\be
P =\pmatrix{ \partial & 0 & 0 \cr
        0 & 0 & \frac{1}{2} \cr
        0 & \frac{1}{2} & 0},
\ee

Using this Hamiltonina operator our equations could be written down 
in 
the hamiltonian form as
\be
\frac{d}{dt}(\phi ,g,h) = P (\frac{\delta H_{3}}{\delta \phi},
\frac{\delta H_{3}}{\delta g},\frac{\delta H_{3}}{\delta h})^{t},
\ee
where $t$ denotes transposition. 

\section{Reduction} 

Using formula (15) it is possible to construct a whole hierarchy of 
integrable equations. In the next we consider the case in which 
$g=h$. 
Then the superbosonic Lax operator $L_{b}$ reduces to the usual Lax 
operator 
for the supersymmetric $N=2, a=4$ Korteweg - de Vries equation. For 
the 
superfermionic case we obtain interesting possibilities in which 

\be
L_{f}=D_{1}D_{2} + \phi +2h\partial^{-1}D_{1}D_{2}h
\ee

This Lax operator produces the new hierarchy of equations also. The 
second 
flow could be obtained from the formulas (21 - 23) and reads as
\bea
\phi_{t} &=& \Big ( -(D_{1}D_{2}\phi -2\phi^{2} +4h_{x}h -
2(D_{1}h)^{2} -
2(D_{2}h)^{2}\Big )_{x}/2 \ \\
h_{t} &=& (-2(D_{1}D_{2}h_{x} -4h_{x}\phi -(D_{1}\phi)(D_{2}h) - 
(D_{2}\phi )(D_{2}h) -h\phi_{x}\Big )/4.
\eea

The third flow produces us the generalization of the supersymmetric 
$N=2,a=4$ KdV equation in the form .
\bea
\phi_{t} &=& \Big ( -\phi_{xx} + 6\phi (D_{1}D_{2}\phi ) - 
3(D_{2}\phi)(D_{1}\phi) + 4\phi^{3} + 12h(D_{1}D_{2}h_{x}) - \cr
&& 12h_{x}(D_{1}D_{2}h) -24h_{x}h\phi +12\phi (D_{1}h)^{2} + 
12\phi (D_{2}h)^{2}\Big )_{x}/4, \ \\
h_{t} &=& \Big ( -4h_{xxx} +12 \phi (D_{1}D_{2}h_{x}) +
6\phi_{x}(D_{1}D_{2}h) - 3(D_{2}\phi_{x})(D_{1}h) \cr
&& -6(D_{2}\phi )(D_{1}h_{x}) +3(D_{1}\phi_{x})(D_{2}h) + 
6(D_{1}\phi )(D_{2}h_{x})+  \cr
&& 12h_{x}((D_{1}h)^{2}+(D_{2}h)^{2}+\phi^{2}) +
6\phi((D_{1}\phi)(D_{1}h) + (D_{2}\phi)(D_{2}h) + \phi_{x}h)\cr
&& 6h_{x}(D_{1}D_{2}\phi) + 3h(D_{1}D_{2}\phi_{x})\Big )/2 .
\eea
These equations are hamiltonian equations with the following 
Hamiltonian 
operator
\be
P=\pmatrix{ \partial & 0 \cr
    0 & \frac{1}{4} },
\ee
\newpage

The hamiltonian which produce the equations (34-35) is

\bea
H_{3} &=& \int dX  \Big ( \phi (D_{1}D_{2}\phi ) + \frac{3}{4}\phi^{3}
-6(D_{1}D_{2}h_{x})h - 6h_{x}h\phi + \cr
&&  + 3\phi (D_{1}h)^{2}+ 3\phi (D_{2}h)^{2} )/12.
\eea
while this which give us  equations (36-37) is
\bea
H_{5} &=& \int dX \Big ( \frac{1}{2}\phi^{2}_{x} 
+\frac{3}{2}\phi^{2}(D_{1}D_{2}\phi ) +\phi^{4} -2ff_{xxx} 
+ 3ff_{x}((D_{1}f)^2+(D_{2}f)^{2}) \cr
&& +\phi (12f(D_{1}D_{2}f_{x}) -12f_{x}(D_{1}D_{2}f) -12\phi f_{x}f + 
6\phi (D_{1}f)^{2} + 6\phi (D_{2}f)^{2}) \Big ).
\eea

Now let us investigate the bosonic limit of the equations (34-37). 
Assuming that 
\be
\phi = \phi_{o} + \theta_{2}\theta_{1}\phi_{1},
\ee
\be
h=\theta_{1}h_{o} + \theta_{2}h_{1}.
\ee
the bosonic sector of the equations 34-35 has the following form
\bea
\phi_{ot} &=& \Big ( -\phi_{1} -2\phi^{2}_{o}-2h^{2}_{o}-2h^{2}_{1} 
\Big 
)_{x}/2, \cr
\phi_{1t} &=& \Big ( \phi_{oxx} -4\phi_{1}\phi_{o} +8h_{o}h_{1x} - 
8h_{1}h_{ox} \Big )_{x}/2, \cr
h_{ot} &=& \Big ( -2h_{1xx} -2\phi_{ox}h_{o} -4\phi_{o}h_{ox} 
-\phi_{1}h_{1}\Big ) /2, \cr
h_{1t} &=& \Big ( 2h_{oxx} -2\phi_{ox}h_{1} - 4\phi_{o}h_{1x} +
\phi_{1}h_{o} \Big )/2.
\eea
The bosonic sector of the equations (36-37) is

\bea
\phi_{ot} &=& \Big ( -\phi_{oxx} +6\phi_{1}\phi_{o} +4\phi_{o}^{3} +
12\phi_{o}(h_{1}^{2} + h_{2}^{2}) \Big )_{x}/4, \ \\
\phi_{1t} &=& \Big ( -\phi_{1xx} -3\phi_{ox}^{2} -6\phi_{oxx}\phi_{o} 
+
12\phi_{1}\phi_{o}^{2} +3\phi_{1}^{2} \cr
&& 12\phi_{1}(h_{1}^2+h_{2}^{2}) + 12h_{oxx}h_{o} -12h_{ox}^{2} -
48\phi_{o}h_{1x}h_{o} + \cr
&& 12h_{1xx}h_{1} -12h_{1x}^{2} +48\phi_{o}h_{1}h_{ox}\Big )_{x}/4,\  
\\
h_{ot} &=& \Big ( -4h_{oxxx} +12h_{ox}h_{o}^{2} +12h_{ox}h_{1}^{2} +
12\phi_{ox}h_{1x} +3\phi_{oxx}h_{1}+ \cr
&& 12\phi_{o}h_{1xx} +12\phi_{o}^{2}h_{ox} 
+12\phi_{ox}\phi_{o}h_{o} +6\phi_{1}\phi_{o}h_{1} \Big )/2, \ \\
h_{1t} &=& \Big ( -4h_{1xxx} +12h_{1x}h_{1}^{2} +12h_{1x}h_{o}^{2} -
12\phi_{ox}h_{ox} -3\phi_{oxx}h_{o}- \cr
&& 12\phi_{o}h_{oxx} +12\phi_{o}^{2}h_{1x} +12\phi_{ox}\phi_{o}h_{1} -
6\phi_{1}\phi_{o}h_{o} \Big )/2. 
\eea
The Hamiltonian operator for these systems could be extracted also 
from the 
supersymmetric Hamiltonian operator (38) and reads as
\bea
P &=& \pmatrix{ 0 & \partial & 0 & 0 \cr
      \partial & 0 & 0 & 0 \cr
     0 & 0 & 0 & 1 \cr
    0 & 0 & -1 & 0 }.
\eea
Similarly we can computed the corresponding hamiltonians using the 
formulas 
(39-40)

Interestingly the system of equations (44-47) allows us to make 
additional 
reduction in which $\phi_{o}=0$. In this limit these equations 
reduces to 
\newpage

\bea
\phi_{1t} &=& \Big ( -\phi_{1xx} +3\phi_{1}^{2} 
+12\phi_{1}(h_{0}^{2}+h_{1}^{2}) + 12h_{oxx}h_{o} \cr 
&&-12h_{ox}^{2} + 12h_{1xx}h_{1}-12h_{1x}^{2} \Big )/4,\ \\
h_{ot} &=& \Big ( -2h_{oxxx} + 6h_{ox}h_{o}^{2} + 6h_{ox}h_{1}^{2} 
\Big ), 
\ \\
h_{1t} &=& \Big ( -2h_{1xxx} + 6h_{1x}h_{1}^{2} + 6h_{1x}h_{o}^{2} 
\Big ).
\eea
and describes the interactions of Korteweg - de Vries field with two 
Modified Korteweg - de Vries fields.

\end{document}